\documentclass{article}
\usepackage[utf8]{inputenc}
\usepackage{amsmath}
\usepackage{graphicx}
\usepackage{hyperref}

\newtheorem{definition}{Definition}

\newcommand{\AP}{\text{AP}}
\newcommand{\AUC}{\text{AUC}}
\newcommand{\NDCG}{\text{NDCG}}

\newcommand{\RECALL}{\text{Recall}}
\newcommand{\PRECISION}{\text{Prec}}
\newcommand{\bx}{\mathbf{x}}

\title{Evaluation Metrics for Item Recommendation under Sampling}
\author{
  \and
  Steffen Rendle\thanks{Google Research, Mountain View, USA}\\
  \texttt{srendle@google.com}
}
\date{}

\begin{document}

\maketitle

\begin{abstract}
The task of item recommendation requires ranking a large catalogue of items given a context.
Item recommendation algorithms are evaluated using ranking metrics that depend on the positions of relevant items.
To speed up the computation of metrics, recent work often uses sampled metrics where only a smaller set of random items and the relevant items are ranked.
This paper investigates sampled metrics in more detail and shows that sampled metrics are inconsistent with their exact version.
Sampled metrics do not persist relative statements, e.g., \emph{algorithm A is better than B}, not even in expectation.
Moreover the smaller the sampling size, the less difference between metrics, and for very small sampling size, all metrics collapse to the AUC metric.
\end{abstract}

\section{Introduction}

Over recent years, item recommendation from implicit feedback has received a lot of attention from the recommender system research community.
At its core item recommendation is a retrieval task, where given a context a catalogue of items should be ranked and the top scoring ones are shown to the user.
Usually the catalogue of items to retrieve from is large: in academic studies usually tens of thousands and industrial applications often many millions.
Finding matching items from this large pool is challenging as the user will usually only explore a few of the highest ranked ones.
For evaluating recommender algorithms, usually sharp metrics such as precision or recall over the few highest scoring items (e.g, top 10) are chosen.
Another popular class are smooth metrics such as average precision or normalized discounted cumulative gain (NDCG) which place a strong emphasis on the top ranked items.

Recently, it has become popular in research papers to speed up evaluation by sampling a small set of irrelevant items and ranking the relevant documents only among this smaller set~\cite{he:www2017,ebesu:sigir2018, hu:kdd2018, yang:rs2018, yang:wsdm2018, krichene:iclr2019, wang:aaai2019}.
In this work, the consequences of this approach are studied.
In particular, it is shown that findings from sampled metrics (even in expectation) can be inconsistent with exact metrics.
This means that if an algorithm A outperforms an algorithm B on the sampled metric, it does not imply that A has a better metric than B when the metric is computed exactly.
This is even a problem in expectation; i.e., with unlimited repetitions of the measurement.
Moreover, a sampled metric has different characteristics than its exact counterpart.
In general, the smaller the sampling size, the less differences there are between different metrics and finally all metrics collapse to the area under the ROC curve where positions are linearly discounted.
This is especially a problem because ranking metrics are designed to focus on the top positions.

This analysis suggests that if a study is really interested in metrics that emphasize the top of the list, it should not use sampled metrics.
Secondly, if sampling is used, the reader should be aware that the reported metric has different characteristics than its name implies.

\section{Evaluating Item Recommendation}

This section starts by formalizing the most common evaluation scheme for item recommendation.
Let there be a pool of $n$ items to recommend from.
For a given instance\footnote{E.g., a user, context, or query.} $\bx$, a recommendation algorithm, $A$, returns a ranked list of the $n$ items.
In an evaluation, the positions, $R(A,\bx) \subseteq \{1,\ldots,n\}$, of the withheld relevant items within this ranking are computed -- $R$ will also be referred to as the \emph{predicted ranks}.
For example, $R(A,\bx) = \{3, 5\}$ means for an instance $\bx$ the algorithm $A$ ranked two relevant items at position $3$ and $5$.
Then, a metric $M$ is used to translate the positions into a single number measuring the quality of the ranking.
This process is repeated for a set of instances, $D=\{\bx_1, \bx_2, \ldots\}$, and an average metric is reported:
\begin{align}
    \frac{1}{|D|} \sum_{\bx \in D} M(R(A,\bx)) .
\end{align}
This problem definition assumes that in the ground truth, all relevant videos are equally preferred by the user, i.e., that the relevant videos are a \emph{set}.
This is the most commonly used evaluation scheme in recommender systems.
In more complex cases, the ground truth includes preferences among the relevant videos.
For example, the ground truth can be a ranked list or weighted set.
The proposed work shows issues with sampling in the simpler setup, which implies that the issues carry over to the more complex case.

In the next sections, first common ranking metrics are formulated as a functions of the predicted rank.
Then sampled metrics, which are often used to speed up computation of metrics, are investigated.
A severe issue with sampled metrics is identified, showing that most sampled metrics are inconsistent and do not even persist qualitative ordering.

\section{Metrics}

This section recaps commonly used metrics for measuring the quality of a ranking.
For convenience, the arguments, $A,\bx$, from $R(A,\bx)$ are omitted whenever the particular algorithm, $A$, or instance, $\bx$, is clear from context.
Instead, the shorter form $R$ is used.

Area under the ROC curve (AUC) measures the likelihood that a random relevant item is ranked higher than a random irrelevant item.
\begin{align}
    \AUC(R)_n =& \frac{1}{|R|(n-|R|)}\sum_{r\in R}\sum_{r' \in (\{1,\ldots, n\} \setminus R)} \delta(r < r')\\
    =& \frac{1}{|R|(n-|R|)} \left[ \sum_{r\in R} (n - r) - \frac{|R|(|R|-1)}{2}\right]\\
    =& \frac{n - \frac{|R|-1}{2} - \frac{1}{|R|}\sum_{r \in R} r}{n-|R|}
\end{align}
with the indicator function $\delta(b) = 1$ iff $b$ is true and $0$ otherwise.
Precision at position $k$ measures the fraction of relevant items among the top $k$ predicted items:
\begin{align}
    \PRECISION(R)_k = \frac{|\{r \in R: r \leq k\}|}{k}
\end{align}
Recall\footnote{Equivalent to `Hit Ratio'.} at position $k$ measures the fraction of all relevant items that were recovered in the top $k$:
\begin{align}
    \RECALL(R)_k = \frac{|\{r \in R: r \leq k\}|}{|R|}
\end{align}
Average Precision at $k$ measures the precision at all ranks that hold a relevant item:
\begin{align}
    \AP(R)_k = \frac{1}{\min(|R|,k)}\sum_{i=1}^k \delta(i \in R) \PRECISION(R)_i
\end{align}
Normalized discounted cumulative gain (NDCG) at $k$ places an inverse log reward on all positions that hold a relevant item:
\begin{align}
    \NDCG(R)_k = \frac{1}{\sum_{i=1}^{\min(|R|,k)} \frac{1}{\log_2(i+1)}} \sum_{i=1}^k \delta(i \in R) \frac{1}{\log_2(i+1)}
\end{align}

\subsection{Simplified Metrics}
\label{sec:simplified_metrics}
The remainder of the paper analyzes these metrics for $|R|=1$, i.e., there exists exactly one relevant item which is ranked at position $r$.
This will simplify the analysis and gives a better understanding about the differences of these metrics.
The metrics of the previous section simplify to:
\begin{align}
    \AUC(r)_n &= \frac{n - r}{n - 1} \\ 
    \PRECISION(r)_k &= \delta(r \leq k) \frac{1}{k} \\
    \RECALL(r)_k &= \delta(r \leq k) \\
    \AP(r)_k &= \delta(r \leq k) \frac{1}{r} \\ 
    \NDCG(r)_k &= \delta(r \leq k) \frac{1}{\log_2(r+1)}
\end{align}
For metrics such as Average Precision and NDCG, it makes sense to define also their unbounded counterpart, e.g., for $k = n$:
\begin{align}
    \AP(r) &= \frac{1}{r} \\
    \NDCG(r) &= \frac{1}{\log_2(r+1)}
\end{align}
Some other popular metrics can be reduced to these definitions:
\emph{Reciprocal Rank} for $|R|=1$ is equivalent to Average Precision, and \emph{Accuracy} is equivalent to Recall at 1, and Precision at 1.

\begin{figure}
    \centering
    \includegraphics[width=0.49\textwidth]{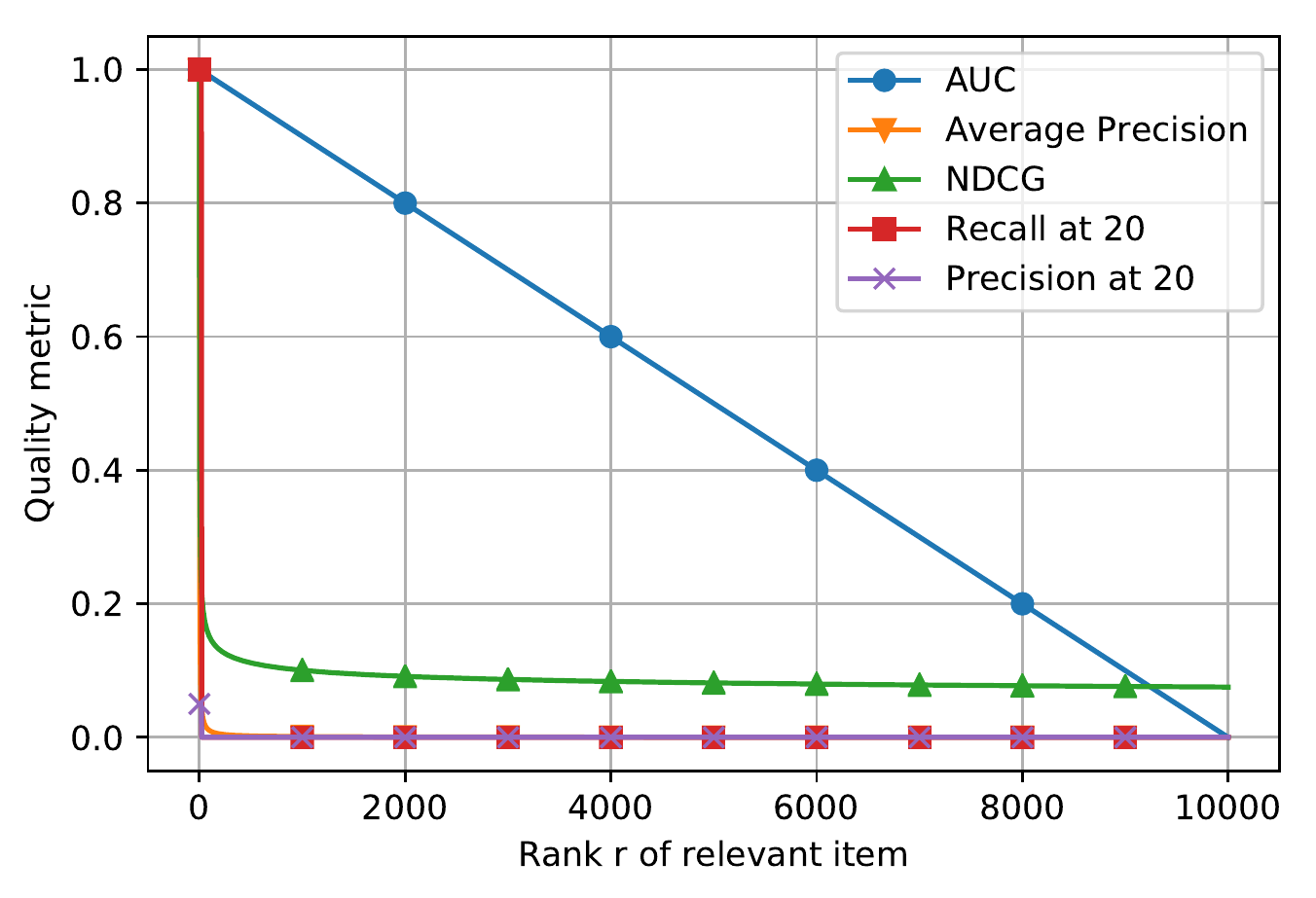}
    \includegraphics[width=0.49\textwidth]{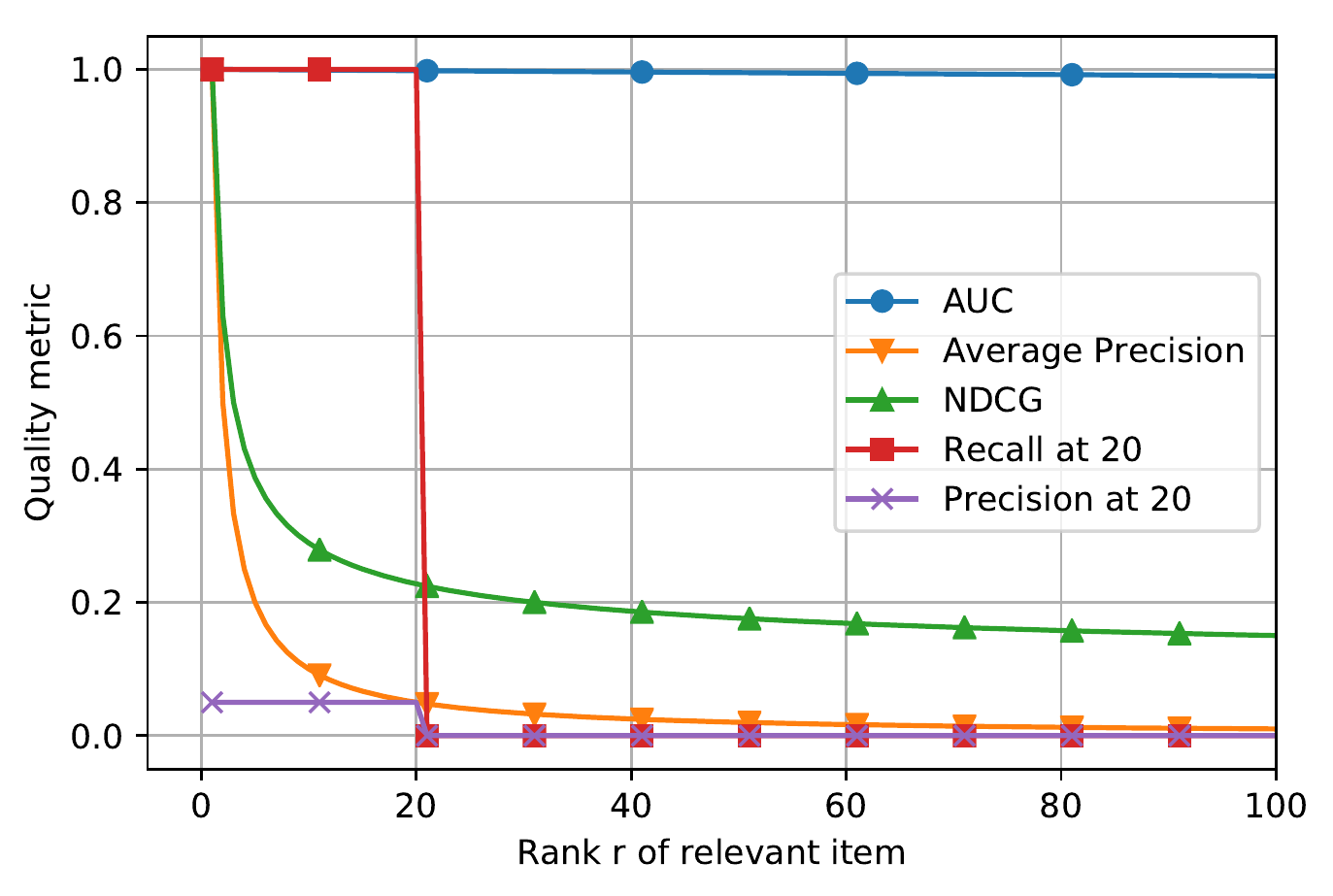}
    \caption{Visualization of metric vs. predicted rank for $n=10,000$. 
    The left side shows the metrics over the whole set of $10,000$ items.
    The right side zooms into the contributions of the top $100$ ranks.
    All metrics besides AUC are top heavy and almost completely ignore the tail.
    This is usually a desired property for evaluating ranking measures because users are unlikely to explore items further down the result list.}
    \label{fig:metrics}
\end{figure}

Figure~\ref{fig:metrics} visualizes how the different ranking metrics trade off the position vs. quality score.
Average precision places the highest importance on the rank, e.g., rank 1 is twice as valuable as rank 2, whereas for NDCG, rank 1 is 1.58 more valuable than rank 2.
The least position aware metric is AUC which places a linear decay on the rank.
E.g., pushing an item from position 101 to 100 is as valuable as pushing an item from position 2 to 1.

\subsection{Example}

\label{sec:metrics_example}

This section concludes with a short example that will be used throughout this work.
Let there be three algorithms $A$, $B$, $C$ and a set of $n=10,000$ items.
Each algorithm is evaluated on five instances (i.e., $|D| = 5$) with one relevant item each.
For each instance, each algorithm creates a ranking and the position at which the relevant item appears is recorded.
Assume that algorithm $C$ manages to rank the relevant item in one of the evaluation instances on position 2, besides this it never achieves a good rank for the other four instances.
Assume algorithm $B$ could rank relevant items in two evaluation instances at position 40.
And algorithm $A$ is never good nor horrible and the relevant items are ranked at position $100$ in each of the five evaluation instances.
The following table shows more details about the predicted ranks and the evaluation metrics that follow from this rank:

\begin{center}
\begin{tabular}{l|l|r|r|r|r}
 & Predicted Ranks          & $\AUC$ &$\AP$ & $\NDCG$ & $\RECALL @10$\\
\hline
A         & 100, 100, 100, 100, 100  & \textbf{0.990} & 0.010 & 0.150 & 0.000\\
B         & 40, 40, 8437, 9266, 4482 & 0.555 & 0.010 & 0.122 & 0.000\\
C         & 212, 2, 743, 5342, 1548  & 0.843 & \textbf{0.101} & \textbf{0.208}& \textbf{0.200}\\
\hline
\end{tabular}
\end{center}
On AUC, algorithm $A$ is clearly the best as it does care about all ranks equally.
For the top heavy metrics (AP, NDCG and Recall), algorithm $C$ is most successful.
This example will be revisited in Section~\ref{sec:sampled_metrics_example} when sampled metrics are discussed.

\section{Sampled Metrics}

Ranking all items is expensive when the number of items, $n$, is large.
Recently, it has become popular to sample a small set of $m$ irrelevant items, add the relevant items, and compute the metrics only on the ranking among this subset~\cite{he:www2017,ebesu:sigir2018, hu:kdd2018, yang:rs2018, yang:wsdm2018, krichene:iclr2019, wang:aaai2019}.
It is common to pick the number of sampled irrelevant items, $m$, in the order of a hundred while the number of items $n$ is much larger, e.g., $m=100$ samples for datasets with $n=\{4k, 10k, 17k, 140k, 2M\}$ items~\cite{he:www2017,ebesu:sigir2018,wang:aaai2019}, $m=50$ samples for $n \in \{2k, 18k, 14k\}$ items~\cite{hu:kdd2018}, or $m=200$ samples for $n \in \{17k, 450k\}$ items~\cite{yang:rs2018}.
This section will highlight that this approach is problematic.
In particular, results can become inconsistent with the exact metrics.

\subsection{Formalization}

Let $\tilde{R}$ be the ranks of the relevant items among the union of relevant items and a sample of $m$ randomly drawn, irrelevant items
It is important to note that $\tilde{R}$ is a random variable, i.e., it depends on the random sample of irrelevant items.
The properties of $\tilde{R}$ will be analyzed in Section~\ref{sec:shape_of_r}.

\subsection{Inconsistency of Sampled Metrics}

A central goal of evaluation metrics is to make comparisons between algorithms, such as, \emph{algorithm $A$ has a higher value than $B$ on metric $M$}.
When comparing algorithms among sampled metrics, we would hope that at least the relative order is preserved in expectation.
This property can be formalized as:
\begin{definition}[Consistency]
A metric $M$ is \underline{consistent} under sampling if the relative order of two algorithms $A$ and $B$ is preserved in expectation.
\begin{align}
    &\frac{1}{|D|} \sum_{\bx \in D} M(R(A,\bx)) > \frac{1}{|D|} \sum_{\bx \in D} M(R(B,\bx)) \notag \\
    \iff&
    E\left[\frac{1}{|D|} \sum_{\bx \in D} M(\tilde{R}(A,\bx))\right] > E\left[\frac{1}{|D|} \sum_{\bx \in D} M(\tilde{R}(B,\bx))\right] \label{eq:consistent}
\end{align}
\end{definition}
The expected value can be simplified:
\begin{align}
    E\left[\frac{1}{|D|} \sum_{\bx \in D} M(\tilde{R}(A,\bx))\right] = \frac{1}{|D|} \sum_{\bx \in D} E\left[M(\tilde{R}(A,\bx))\right] 
\end{align}
If a metric is inconsistent, then measuring $M$ on a subsample is not a good indicator of the performance of M.

\subsection{Example}
\label{sec:sampled_metrics_example}

Now, the example from Section~\ref{sec:metrics_example} is revisited and the same measures are computed using sampling.
Specifically, $m=99$ random irrelevant items are sampled, the position $\tilde{r}$ of the relevant item among this sampled subset is found, and then the metrics are computed for the rank $\tilde{r}$ within the subsample.
This procedure with a comparable small sample size is commonly used in recent work~\cite{he:www2017,ebesu:sigir2018,hu:kdd2018,yang:rs2018,wang:aaai2019}.

The following table shows the sampled metrics for the example from Section~\ref{sec:metrics_example}.
As this is a random process, for better understanding of its outcome, here it is repeated 1000 times and the average and standard deviation is computed\footnote{In a real evaluation, the process would not be repeated because this would contradict the motivation of sampling to save time.}.
\begin{center}
{\small
{\setlength{\tabcolsep}{0.44em}
\begin{tabular}{l|l|r|r|r|r}
 & Predicted Ranks          & $\AUC$ &$\AP$ & $\NDCG$ & $\RECALL @10$\\
 \hline
A         & 100, 100, 100, 100, 100  & $\textbf{0.990} {\scriptstyle \pm0.004}$ & $\textbf{0.630} {\scriptstyle \pm0.129}$ & $\textbf{0.724} {\scriptstyle \pm0.097}$ & $\textbf{1.000} {\scriptstyle \pm 0.000}$ \\
B         & 40, 40, 8437, 9266, 4482 & $0.555 {\scriptstyle \pm0.014}$ & $0.336 {\scriptstyle \pm0.073}$ & $0.444 {\scriptstyle \pm0.054}$ & $0.400 {\scriptstyle \pm0.000}$ \\
C         & 212, 2, 743, 5342, 1548  & $0.843 {\scriptstyle \pm0.014}$ & $0.325 {\scriptstyle \pm0.050}$ & $0.460 {\scriptstyle \pm0.039}$ & $0.567 {\scriptstyle \pm0.092}$\\
\hline
\end{tabular}
}}
\end{center}
Compared to the exact metrics in Section \ref{sec:metrics_example}, even the relative ordering of metrics completely changed.
On the exact metrics, C is clearly the best with a 10x higher average precision than B and A. But it has the lowest average precision when sampled measurements are used. A and B perform the same on the exact metrics, but A has a 2x better average precision on the sampled metrics. Sampled average precision does not give any indication of the true relationship among the methods.
Similarly, sampled NDCG and sampled Recall at 10 do not agree with the exact metrics.
Only AUC is consistent between sampling and exact computation.
The other metrics are inconsistent.

\begin{figure}
    \includegraphics[width=0.49\textwidth]{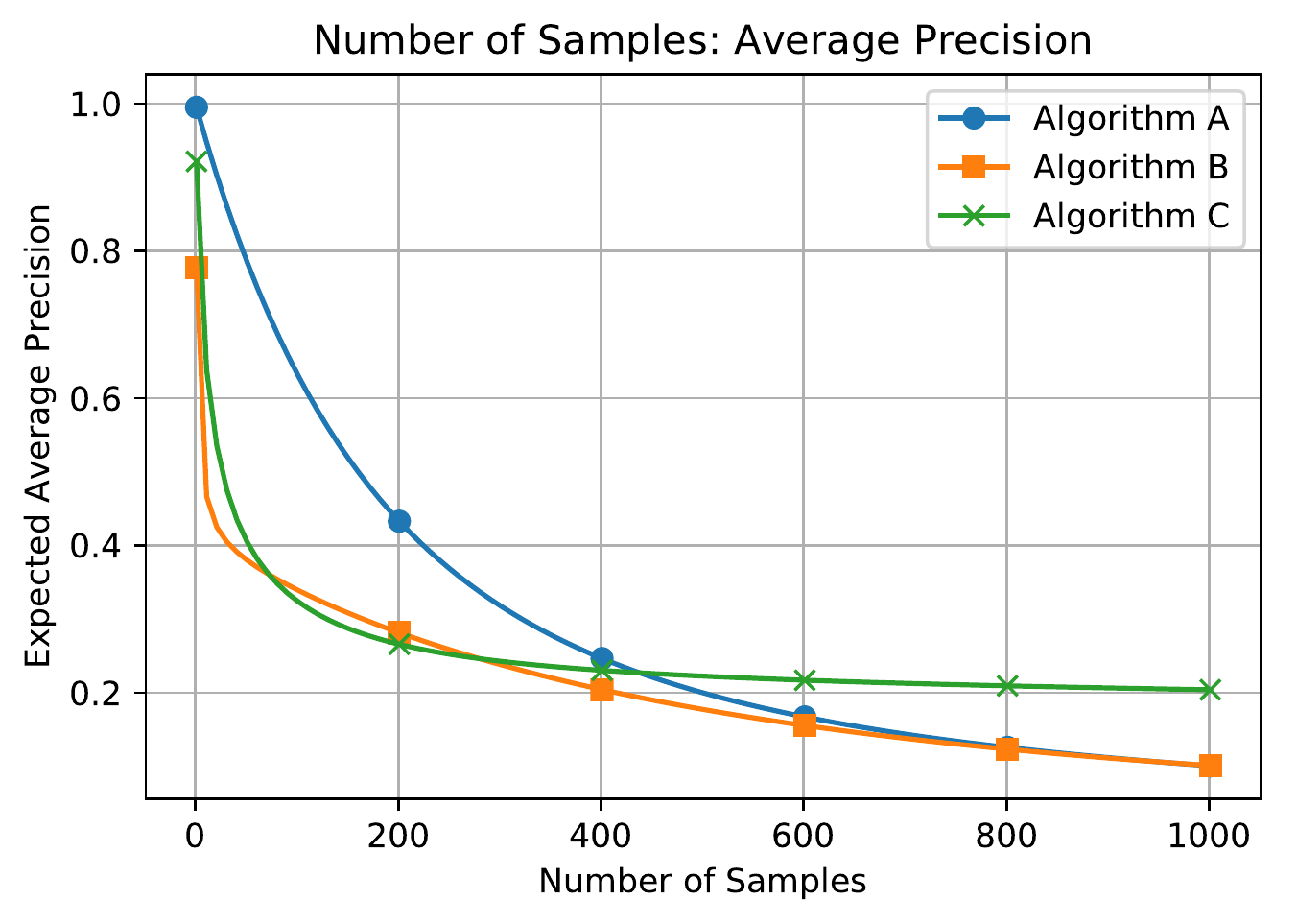}
    \includegraphics[width=0.49\textwidth]{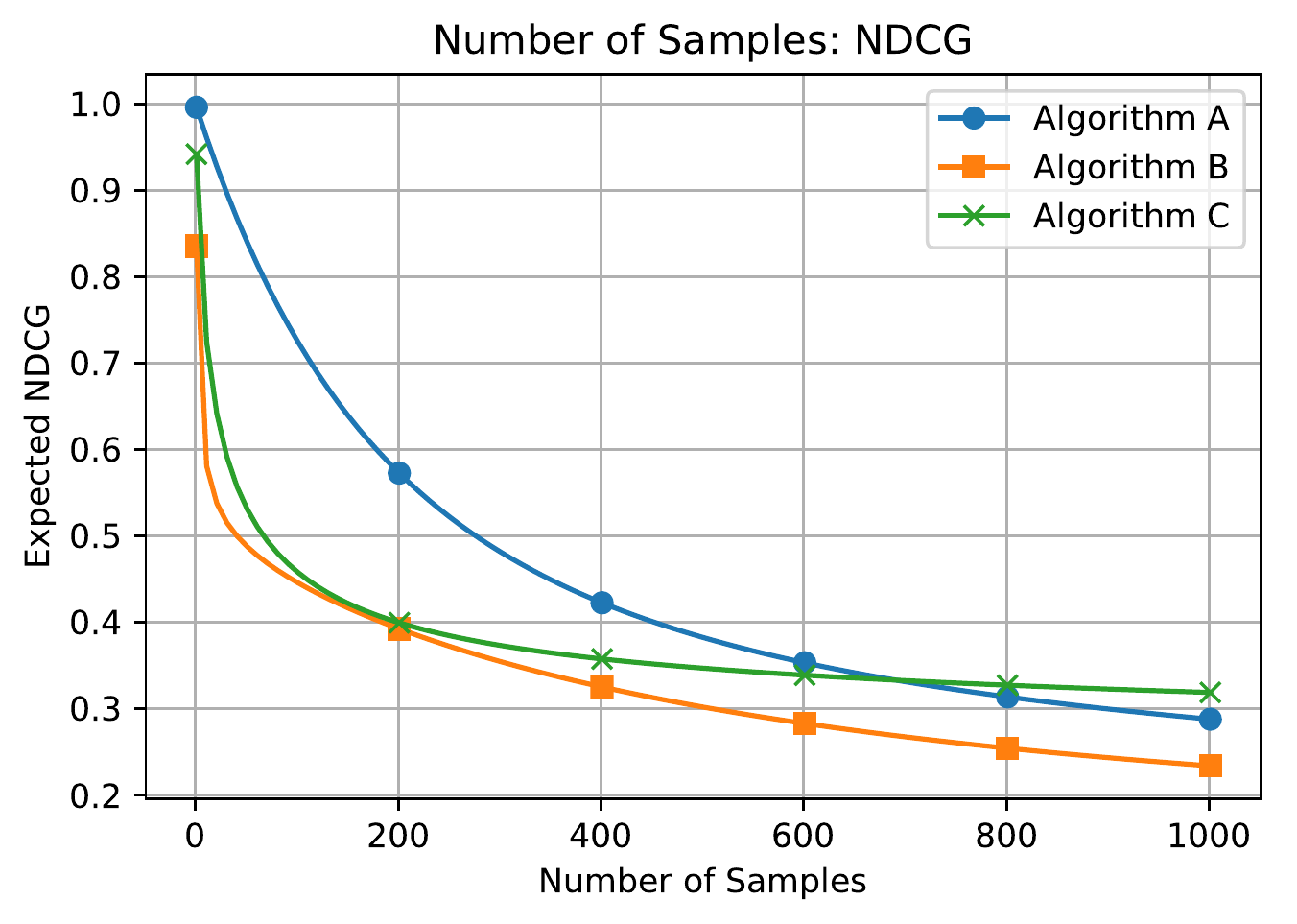}
    \includegraphics[width=0.49\textwidth]{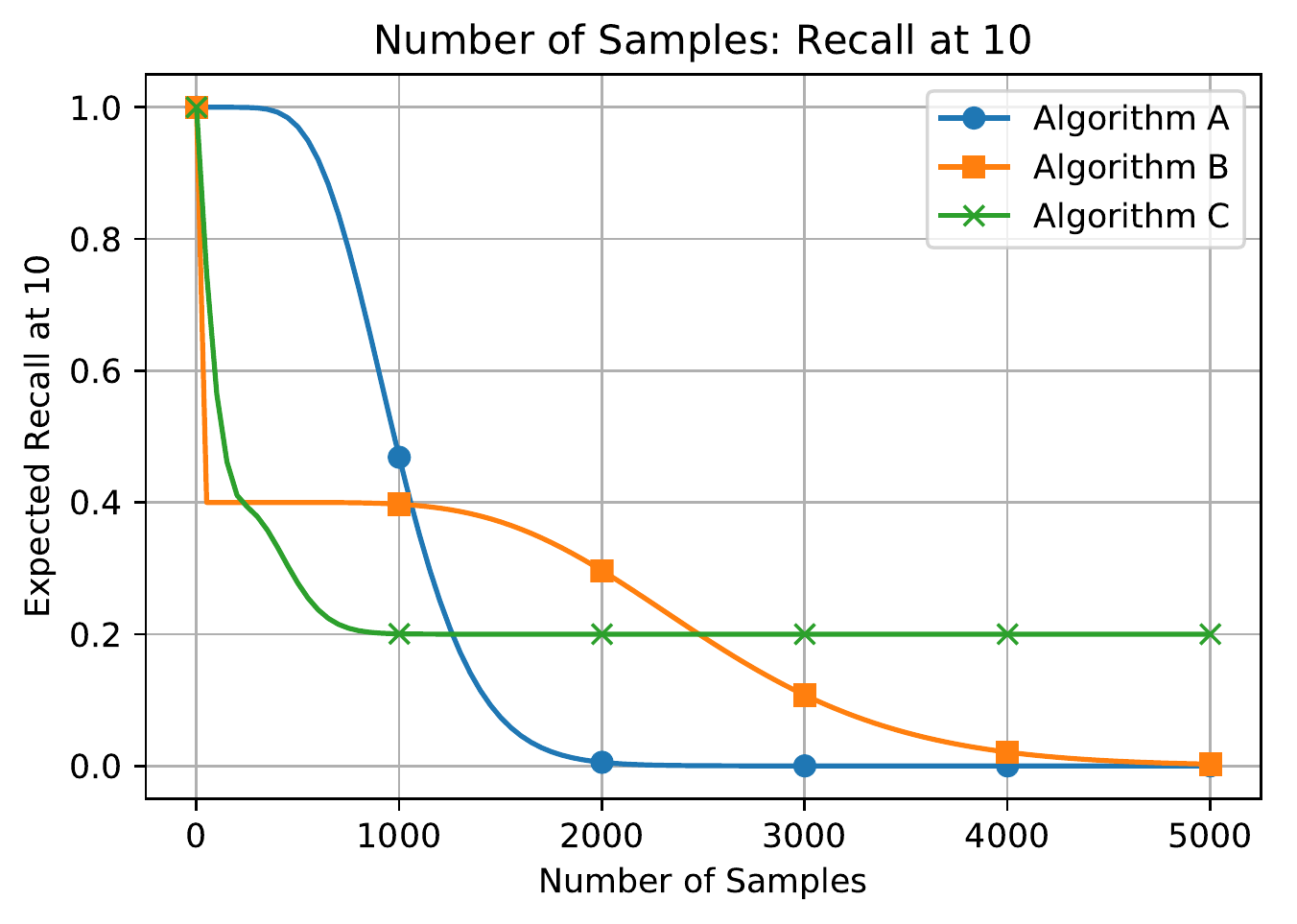}
    \includegraphics[width=0.49\textwidth]{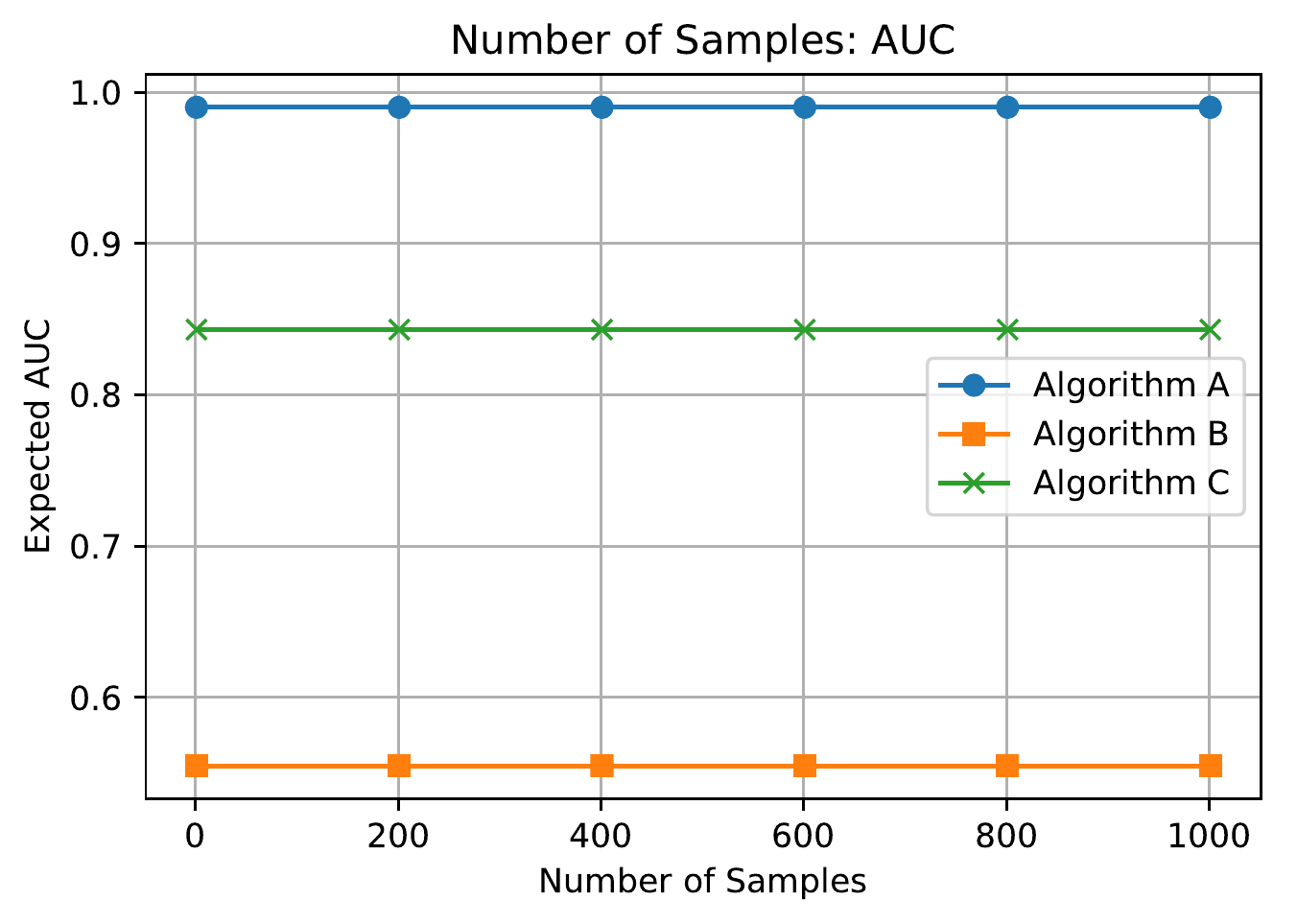}
    \caption{Expected sampling metrics for the running example (Section \ref{sec:metrics_example} and \ref{sec:sampled_metrics_example}) while increasing the number of samples.
    For Average Precision, NDCG and Recall, even the relative order of algorithm performance changes with the number of samples.
    That means, conclusions drawn from a subsample are not consistent with the true performance of the algorithm.}
    \label{fig:example_vary_m}
\end{figure}

Figure~\ref{fig:example_vary_m} shows the same study as in the previous table but now for any choice of number of samples, $m$.
Again, this figure shows the inconsistency of sampled metrics.
Even the relative ordering of algorithms changes with an increasing sample size.
For example, for average precision, depending on the number of samples, any conclusion could be drawn: A better than C better than B (for sample size $< 50$), A better than B better than C (for sample size $\approx 200$), C better than A better than B (for sample size $\approx 500$), and finally C better than A equal B (for large sample sizes).
Similar observations can be made for NDCG.
For recall the dependency on number of samples is even more chaotic and it takes about $m=5,000$ samples out of $n=10,000$ items to get reliable results. 
Only AUC is consistent and no matter which sampling size, the expected metrics are consistent.

\subsection{Rank Distribution under Sampling}
\label{sec:shape_of_r}

This section takes a closer look at the sampling process and derives equations for the distribution of ranks, $\tilde{R}$ and the expected metrics.
For simplicity, the analysis is restricted to rankings with exactly one relevant item, i.e., $|\tilde{R}| = 1$.
Again, $\tilde{r}$ is used with the simplified metrics from Section~\ref{sec:simplified_metrics}.

When an irrelevant item is sampled uniformly, it can either rank higher or lower than the relevant item.
If the true rank of the relevant item is $r$ and the number of all items is $n$, then the probability that the sampled item $j$ is ranked above $r$ is:
\begin{align}
    p(j < r) = \frac{r-1}{n-1}
\end{align}
For example, if $r$ is at position $1$, the likelihood of a random irrelevant being ranked higher is $0$.
If $r=n$, then the likelihood is $1$.
Note that the pool of all possible sampled items excludes the truly relevant item and thus has size $n-1$.

Repeating the sampling procedure $m$ times with replacement and counting how often an item is ranked higher, corresponds to a Binomial distribution.
In other words, the rank $\tilde{r}$ obtained from the sampling process follows $\tilde{r} \sim B\left(m, \frac{r-1}{n-1}\right) + 1$.
If there are no successes in getting a higher ranked item, the rank remains $1$, if all $m$ samples are successful, the rank is $m+1$.

The expected value of the metrics under this distribution is
\begin{align}
    E[M_{m+1, k}(\tilde{r})] = \sum_{i=1}^{m+1} p(\tilde{r} = i) M_{m+1, k}(i), \label{eq:expected_metric}
\end{align}
where the metric, $M$, is now measured among $m+1$ items -- one relevant and $m$ sampled ones. 

\begin{figure}
    \includegraphics[width=0.49\textwidth]{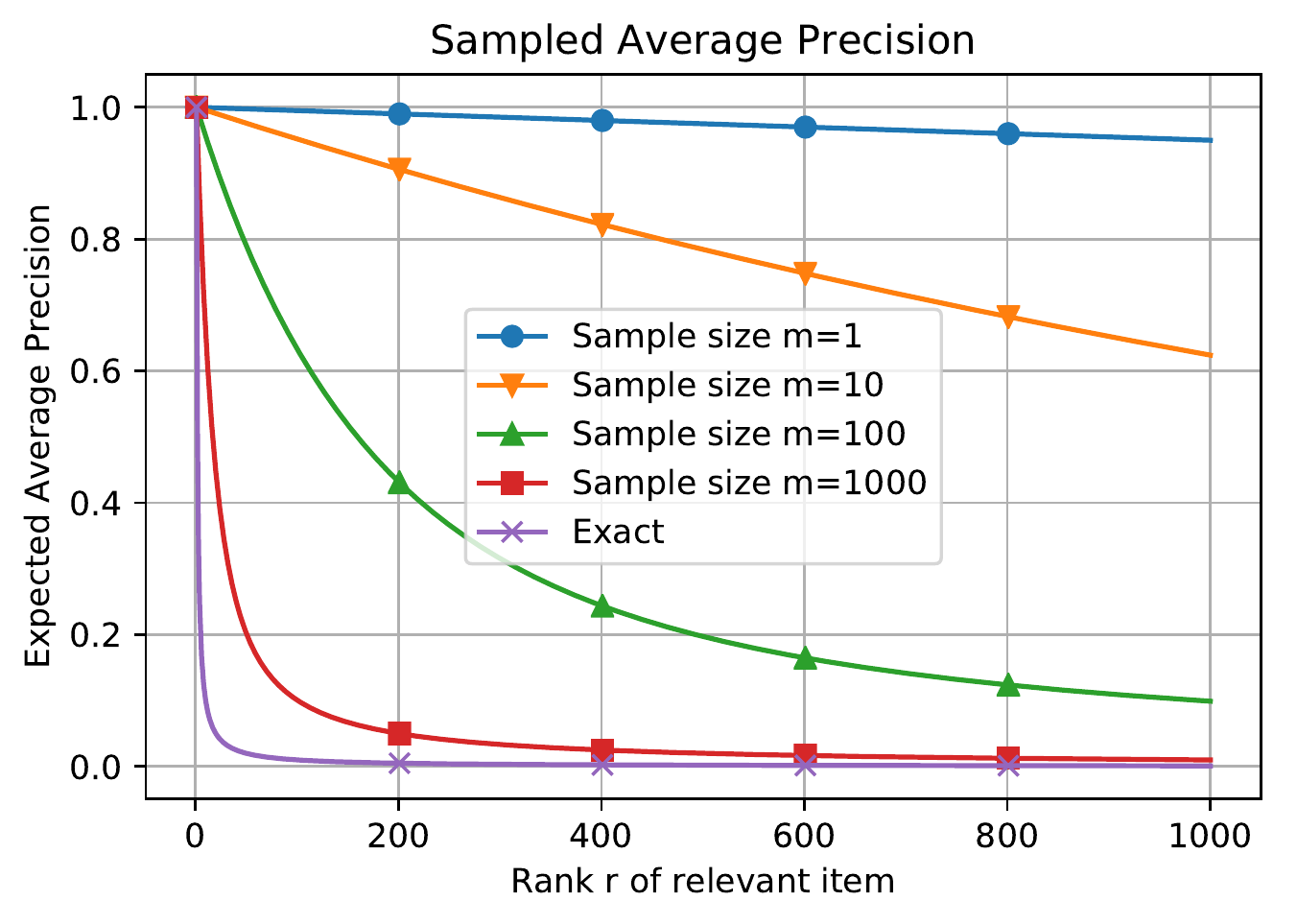}
    \includegraphics[width=0.49\textwidth]{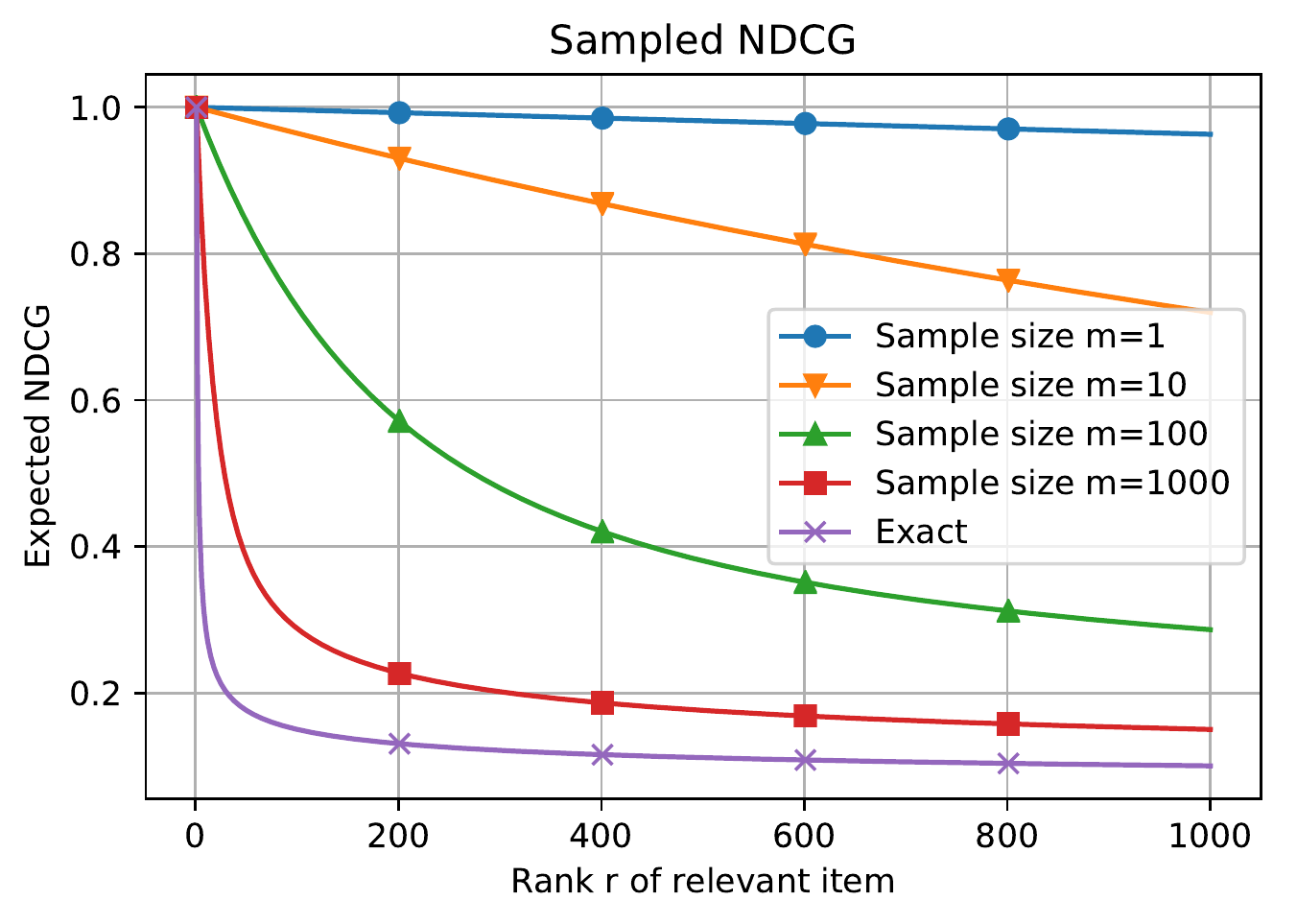}
    \includegraphics[width=0.49\textwidth]{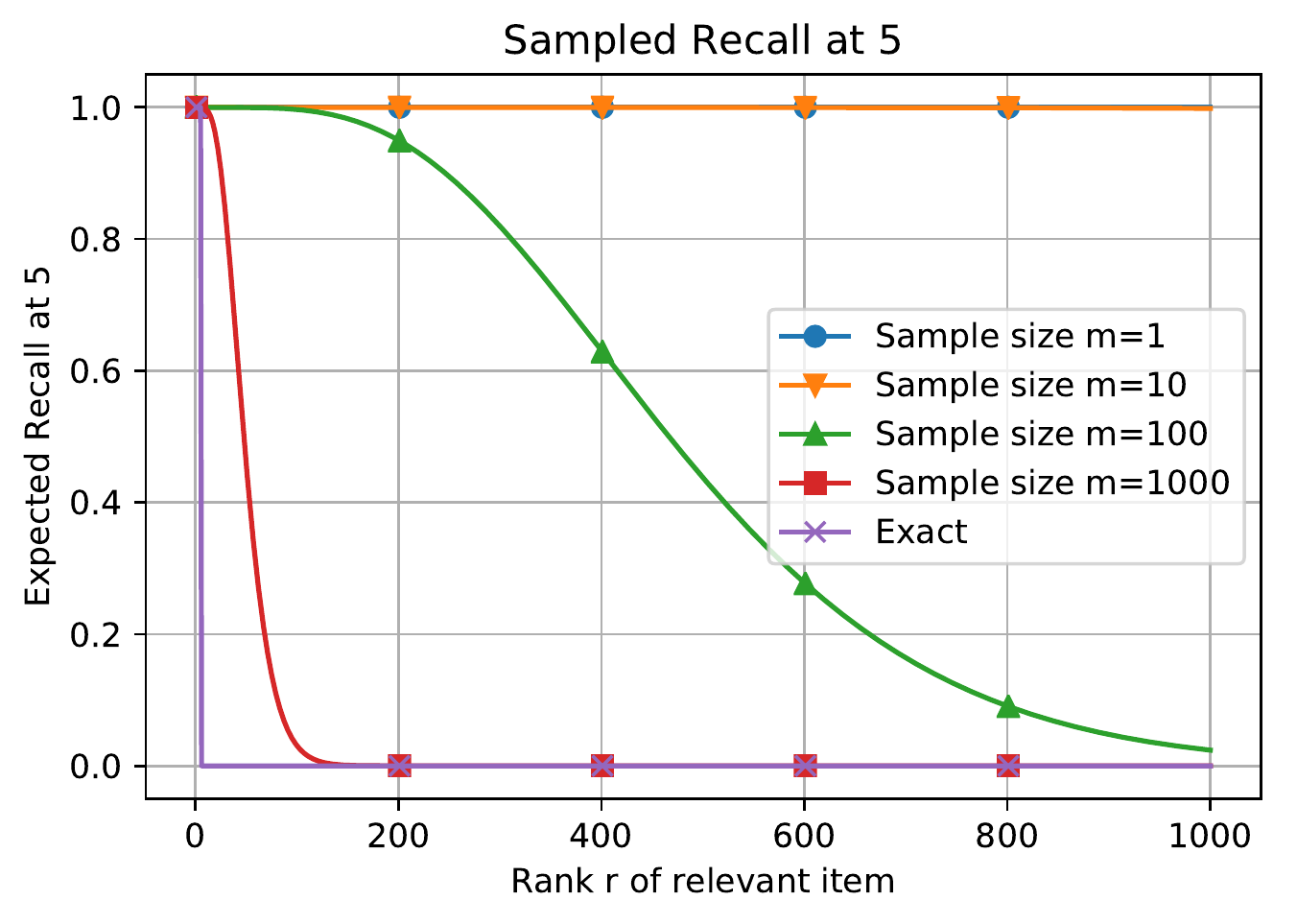}
    \includegraphics[width=0.49\textwidth]{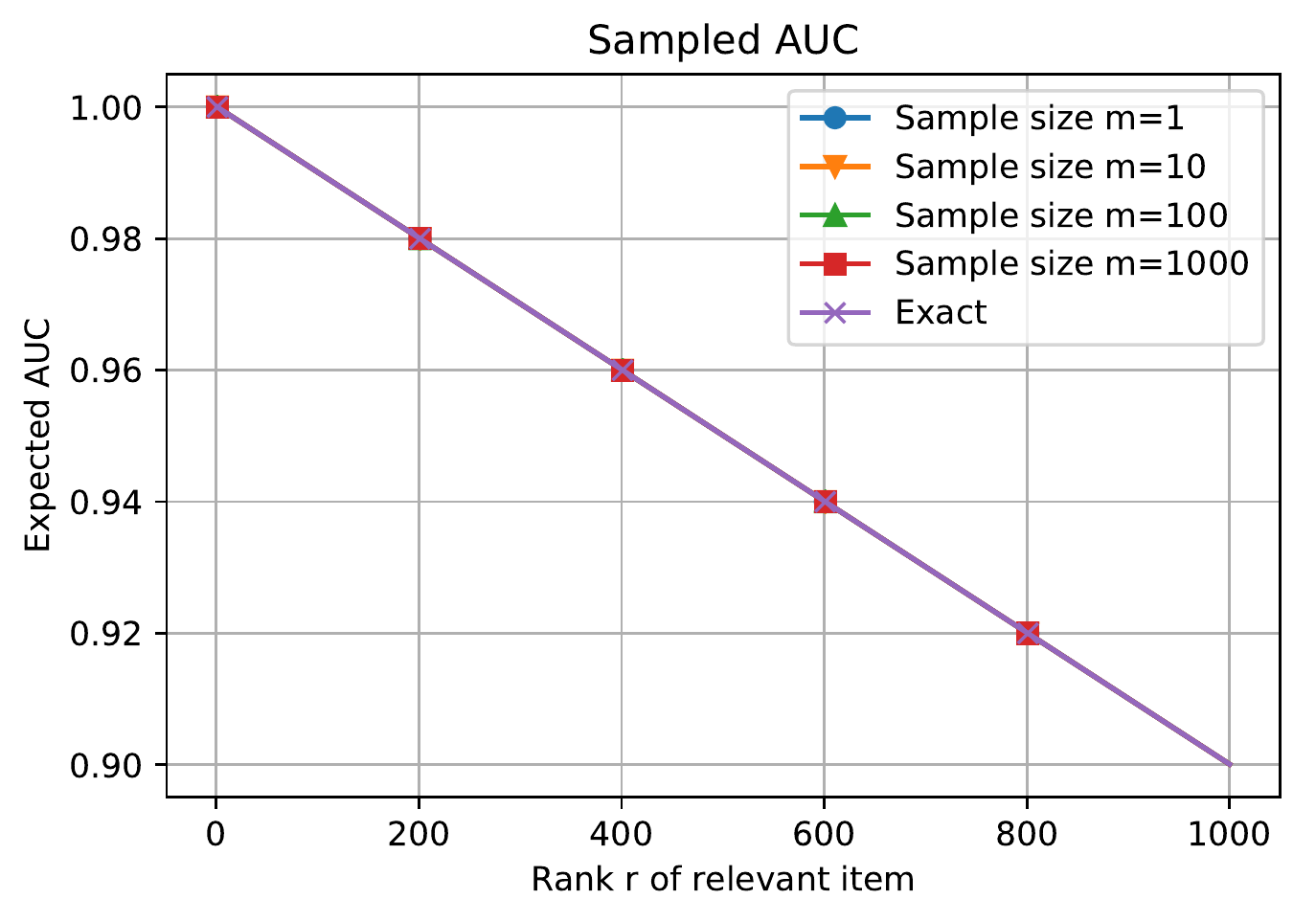}
    \caption{Characteristics of a sampled metric with a varying number of samples.
    Sampled Average Precision, NDCG and Recall change their characteristics substantially compared to exact computation of the metric.
    Even large sampling sizes ($m=1000$ samples of $n=10000$ items) show large deviations.
    Note this plot zooms into the top 1000 ranks out of $n=10000$ items.}
    \label{fig:expected_ranking}
\end{figure}

Figure~\ref{fig:expected_ranking} visualizes the expected metrics computed by sampling.
The figure highlights that sampled metrics have a different shape then exact metrics.
Metrics like Average Precision or NDCG are much less top heavy.
Even sharp metrics such as recall become smooth.
Only AUC remains the same.
In general, all metrics get closer and closer to AUC behavior.
I.e., they become linear and less focused on top ranks.

\subsection{Expected Metrics}

This section analyzes sampled metrics in a more formal way by applying particular metrics to the general equation for expected metrics, eq.~(\ref{eq:expected_metric}). 
The discussion focuses on uniform sampling with replacement, i.e., Binomial distributed ranks.
Similar results hold for uniform sampling without replacement.
In this case, the distribution is hypergeometric, with population size $n-1$, where a pool of $r-1$ items can be potential successes.
When appropriate, this variation will be discussed as well.

\subsubsection{Expected AUC}
\label{sec:expected_auc}

First, AUC is a linear function in the rank:
\begin{align}
    \AUC_n(r) = \frac{n-r}{n-1} = -\frac{1}{n-1}\,r + \frac{n}{n-1} = \text{const}_1\, r + \text{const}_2 \label{eq:auc_is_linear}
\end{align}
That means the expected value of sampled AUC can be simplified to
\begin{align}
    E[\AUC_{m+1}(\tilde{r})] = \AUC_{m+1}(E[\tilde{r}]),
\end{align}
and with the expected value of the Binomial distribution:
\begin{align}
    E[\AUC_{m+1}(\tilde{r})] &= \AUC_{m+1}\left(1 + m \frac{r-1}{n-1}\right) \\
                             &= \frac{m+1 - 1 - m \frac{r-1}{n-1}}{m +1 -1} \\
                             &= \frac{m - m \frac{r-1}{n-1}}{m} = 1 - \frac{r-1}{n-1} \\
                             &= \frac{n- r}{n-1} = \AUC_n(r)
\end{align}
That means AUC measurements created by sampling are unbiased estimators of the exact AUC.
This result is not surprising because the AUC can alternatively be defined as the expectation that a random relevant item is ranked over a random irrelevant item.
Consequently, AUC is a consistent metric under sampling.

This result holds also for any sampling distributions where the expected value of the sampled rank is $1 + m \frac{r-1}{n-1}$.
For example, this is also true for sampling from a hypergeometric distribution -- i.e., uniform sampling without replacement. 

\subsubsection{Cut-off metrics}
For a cutoff metric such as recall or precision:
\begin{align}
    E[\RECALL_k(\tilde{r})]
    &= \sum_{i=1}^{m+1} p(\tilde{r} = i) \RECALL_k(i) = \sum_{i=1}^{m+1} p(\tilde{r} = i) \delta(i \leq k) \\
    &= \sum_{i=1}^{k} p(\tilde{r} = i)  = \text{CDF}\left(k-1; m, \frac{r-1}{n-1}\right)
\end{align}
Clearly, the CDF of a binomial distribution is not a linear function and thus cut-off metrics are inconsistent.
This analysis carries over to any sampling distribution, including the hypergeometric distribution. 

\subsubsection{Average Precision}

For the expected value of sampled average precision, the following analysis assumes that the relevant item is not ranked on the first position, i.e., $r > 1$ and consequently $p(j<r) > 0$ -- otherwise the sampled rank is a constant, $\hat{r} == 1$, and the metric is always 1.
\begin{align}
 E[\AP(\tilde{r})]
    &= \sum_{i=1}^{m+1} p(\tilde{r} = i) \AP(i) = \sum_{i=1}^{m+1} p(\tilde{r} = i) \frac{1}{i}\\
    &= \frac{1-(1-p(j < r))^{m+1}}{p(j < r)\,(m+1)} = \frac{1-\left(\frac{n-r}{n-1}\right)^{m+1}}{(r-1)\frac{m+1}{n+1}}
\end{align}
Interestingly, this can be rephrased as:
\begin{align}
 \frac{1-\AUC_n(r)^{m+1}}{r-1} \, \left(\frac{n+1}{m+1}\right) =  \frac{1-\AUC_n(r)^{m+1}}{r-1} \, \text{const}
\end{align}
If $AUC_n(r)^{m+1} \approx 0.0$, this would be closely related to the unsampled average precision metric.
However, as soon as the relevant item is reasonably highly ranked (i.e., AUC is close to $1.0$), it takes many samples $m$ to approach $0$.

\subsubsection{Small Sampling Size}

The final analysis investigates the case where the number of samples is $m=1$.
This analysis will show the behavior of sampling methods in the limit.
If $m=1$, then for any metric $M$ and sampling distribution:
\begin{align}
    E[M(\tilde{r})] 
    &= p(\tilde{r} = 1) M(1) + p(\tilde{r} = 2) M(2) \\
    &= p(\tilde{r} = 1) M(1) + (1-p(\tilde{r} = 1)) M(2) \\
    &= p(\tilde{r} = 1) [M(1)-M(2)] + M(2)
\end{align}
For uniform sampling\footnote{Here $m=1$, so it does not matter if with or without replacement.} of items, the probability to sample an item that is ranked after $r$, i.e., $\tilde{r}=1$, is $\frac{n-r}{n-1}$.
Now,
\begin{align}
    E[M(\tilde{r})] 
    &= \frac{n-r}{n-1} (M(1)-M(2)) + M(2) \\
    &= r\frac{M(2)-M(1)}{n-1} + \frac{n\,M(1) - M(2)}{n-1} \\
    &= r\,\text{const}_1 + \text{const}_2
\end{align}
No matter which metric is used, the result is a linear function of the true rank, $r$.
If we only care about the ordering produced by two different metrics on a set of rankings (eq. \ref{eq:consistent}), we can ignore $\text{const}_2$.
Similarly, for the first constant, $\text{const}_1$, only the sign matters when comparing two sets of ranking.
This sign of $M(2)-M(1)$ depends on how much ranking a relevant item at position 1 is preferred over ranking it at position 2.
Metrics that cannot distinguish between the first and second position, such as precision and recall at $k \geq 2$, are an exception and in this case the sampled metric is always constant and not useful at all.
For any reasonable metric, $\text{const}_1$ should be negative, i.e., ranking at position 1 gives a higher metric than position 2.
To summarize, for $m=1$ all metrics will give qualitatively the same result.
There is no reason to choose one metric over the other if we are only interested in relative statements such as "metric of $A$ is higher than metric of $B$".

Finally, all of these sampled metrics ($m=1$) give the same qualitative statement (in expectation) as exhaustive $\AUC$.
This can be shown in two ways:
(a) Section~\ref{sec:expected_auc} proofed that sampled $\AUC$ is consistent with exhaustive $\AUC$.
The previous analysis argued that all sampled metrics, including sampled $\AUC$, are indistinguishable for $m=1$.
Consequently all sampled metrics for $m=1$ are consistent with $\AUC$.
(b) Eq.~\ref{eq:auc_is_linear} showed that $\AUC$ is a linear function in the rank itself.

The discussion above shows that it does not make sense to choose different metrics for $m=1$; any sensible metric gives the same qualitative statement.
Extrapolating this observation to larger sampling sizes, gives another indication that sampled metrics blur the differences between metrics.
A similar observation can be found in Figure~\ref{fig:expected_ranking} and \ref{fig:example_vary_m} where all metrics behave similar for small samples sizes.
Consequently, if we are really interested in sharp metrics, sampling should be avoided.

\section{Conclusion}

This work investigated the behavior of evaluation metrics under sampling.
It has shown that most metrics are inconsistent under sampling and can lead to false discoveries.
In particular, under heavy sampling the difference between metrics blurs and metrics are less and less distinguishable.
Moreover, metrics are usually motivated by applications, e.g., does the top 10 list contain a relevant item?
Sampled metrics do not measure the intended quantities -- not even in expectation.
This work suggests that if an experimental study really cares about particular metrics, then sampling should be avoided.
If the characteristic of a metric is not of central importance, the reader should be at least aware that a sampled metric measures a different quantity than the exact metric and especially for small sampling sizes, the conclusions drawn can be contradicting.

\section*{Acknowledgements}

I would like to thank Walid Krichene, Nicolas Mayoraz and Li Zhang for their helpful comments and suggestion.

\end{document}